\documentclass{llncs}

\usepackage{microtype}
\usepackage{xspace} 
\usepackage{amssymb}
\usepackage{amsmath}
\usepackage{stmaryrd}
\usepackage{algorithm,algorithmic}
\usepackage{booktabs} 
\usepackage{multirow}
\usepackage{tikz}
\usepackage{placeins}
\usetikzlibrary{matrix,shapes,arrows,positioning,backgrounds,calc}
\usetikzlibrary{decorations.text}

\newtheorem{observation}[theorem]{Observation}

\newcommand{\etal}{\emph{et al.}\@\xspace}

\definecolor{a}{HTML}{FFB338}
\definecolor{b}{HTML}{FC6039}
\definecolor{c}{HTML}{FF194D}
\definecolor{d}{HTML}{F73CB5}
\definecolor{e}{HTML}{E63DF4}
\definecolor{f}{HTML}{B974FF}
\definecolor{g}{HTML}{887FFF}
\definecolor{h}{HTML}{86AEFE}
\definecolor{i}{HTML}{42C3E9}
\definecolor{j}{HTML}{43E7C7}
\definecolor{k}{HTML}{44E481}
\definecolor{l}{HTML}{4CE246}
\definecolor{m}{HTML}{98E246}

\newcommand{\FUNCTION}[1]{\STATE \textbf{procedure} #1 \begin{ALC@g}}
\newcommand{\ENDFUNCTION}{\end{ALC@g}\STATE \textbf{end procedure}}

\newcommand{\CASE}[1]{\STATE \textbf{case} #1\textbf{:} \begin{ALC@g}}
\newcommand{\ENDCASE}{\end{ALC@g}}

\newcommand{\DEFAULT}{\STATE \textbf{default:} \begin{ALC@g}}
\newcommand{\ENDDEFAULT}{\end{ALC@g}}
\newcommand{\DEFAULTLINE}[1]{\STATE \textbf{default:} }

\title{The gene family-free median of three}

\author{Daniel Doerr\inst{1} \and Pedro Feij\~ao\inst{2} \and Metin Balaban\inst{1} \and Cedric Chauve\inst{3}}
\institute{School of Computer and Communication Sciences, EPFL, 1015 Lausanne,
Switzerland \texttt{daniel.doerr@epfl.ch} \and Faculty of Technology, Bielefeld University, 33615 Bielefeld, Germany \and Department of Mathematics, Simon Fraser University, Burnaby BC, Canada} 

\begin{document}

\maketitle

\begin{abstract}
    The gene family-free framework for comparative genomics aims at 
    developing methods for gene order analysis that do not
    require prior gene family assignment, but work directly on a sequence similarity 
    multipartite graph. We present a model for constructing a median of three genomes in this 
    family-free setting, based on maximizing an objective function that generalizes the classical 
    breakpoint distance by integrating sequence similarity in the score of a gene 
    adjacency. 
    We show that the corresponding computational problem is MAX~SNP-hard and
    we present a \mbox{0-1}~linear program for its exact solution. The result of 
    our FF-median program is a median genome with median genes associated to extant 
    genes, in which median adjacencies are assumed to define positional orthologs.
    We demonstrate through simulations and comparison with the OMA orthology
    database that the herein presented method is able compute accurate medians and 
    positional orthologs for genomes  comparable in size of bacterial genomes.
 \end{abstract}

\section{Introduction} 


The prediction of evolutionary relationships between genomic sequences is a long-standing problem in computational biology.  According to Fitch~\cite{Fitch:2000cch}, two genomic sequences are called \emph{homologous} if they descended from a common ancestral sequence. Furthermore, Fitch identifies different events that give rise to a branching point in the phylogeny of homologous sequences, leading to the concepts of orthologous genes (who descend from their last common ancestor through a speciation) and paralogous genes (descending from their last common ancestor through a speciation), that reach far beyond evolutionary genomics~\cite{Gabaldon:2013cch}. Until quite recently, orthology and paralogy relationships were mostly inferred from sequence similarity. However it is now well accepted that the syntenic context can carry valuable evolutionary information, which has lead to the notion of \textit{positional orthologs}~\cite{Dewey:2011bp}. In the present work, we describe a method to compute groups of likely orthologous genes for a group of three genomes, through a new problem we introduce, the \textit{gene family-free median of three}.

Most methods for detecting potential orthologous groups require a prior clustering of the genes of the considered genomes into \textit{homologous gene families}, defined as groups of genes assumed to originate from a single ancestral gene; clustering protein sequences into families is already in itself a difficult problem.

Here, we follow the matching-based approach, framed within the gene family-free principle, that embodies the idea to perform gene order analysis without the prerequisite of gene family or homology assignments. Instead, we are given all-against-all \emph{gene similarities} through a symmetric and reflexive \emph{similarity measure} $\sigma: \Sigma \times \Sigma \to \mathbb R_{\geq
 0}$ over the universe of genes $\Sigma$~\cite{Braga:2013}. We use sequence similarity but other similarity measures can fit the previous definition. 
Gene family or homology assignments represent a particular subgroup of gene similarity functions that require transitivity. Independent of the particular similarity measure $\sigma$, relations between genes imposed by $\sigma$ are considered as candidates for homology assignments. A gene family-free research program was outlined in~\cite{Braga:2013} (see also~\cite{Doerr:2015phd}) and has so far  been developed for the pairwise comparison of genomes~\cite{Doerr:2012dt,Martinez:2015cch,Kowada:2016cch} and shown to be effective for orthology analysis~\cite{Lechner:2014dt}.

In Section~\ref{sec:ffmedian} we introduce a new genome median problem in the family-free framework, that generalizes the traditional breakpoint median problem~\cite{tannier2009zs}. For a group of three genomes, the input of the family-free median problem is a tripartite similarity graph of pairwise gene similarities. Informally, a median of three is defined as both a set of median genes -- each defined by three extant genes forming a clique in the similarity graph, scored according the edges of this clique --, forming a set of median adjacencies, each  supported by at least one extant gene adjacency (hence any median gene belongs to at least one median adjacency). A median is optimal if it maximizes the sum of the scores of its median genes. Hence, the optimization criterion of this problem fully  integrates both sequence similarity and synteny conservation. In
Section~\ref{sec:complexity} we study its the computational complexity and give
an exact algorithm for its solution. We show that our method can be used for
positional ortholog prediction in simulated and real data sets of bacterial
genomes in Section~\ref{sec:results}. 

\section{The gene family-free median of three}\label{sec:ffmedian}

\paragraph{Extant genomes, genes and adjacencies.}
In this work, a genome $G$ is entirely represented by a tuple $G \equiv
(\mathcal C, \mathcal A)$, where $\mathcal C$ denotes a non-empty set of unique
genes, and $\mathcal A$ is a set of \emph{adjacencies}.
Genes are represented by their \emph{extremities}, i.e.,
a gene $g \equiv (g^{\text{t}}, g^{\text{h}})$, $g \in \mathcal C$, consists of
a \emph{head} $g^{\text{h}}$ and a \emph{tail} $g^{\text{t}}$. Telomeres are modeled explicitly,
as special genes of $\mathcal C(G)$ with a single extremity, denoted by
``$\circ$''. Extremities $g^a, \bar g^b$, $a,b \in \{\text{h}, \text{t}\}$ of
any two genes $g, \bar g$ can form an adjacency. 
In the following, we will conveniently use the notation $\mathcal C(G)$ and
$\mathcal A(G)$ to denote the set of genes and the set of adjacencies of genome
$G$, respectively. We indicate the presence of an adjacency $\{x^a_1, x_2^b\}$ in an
extant genome $X$ by
\begin{align} \label{eqn:indicator}
    \mathbb I_X(x_1^a, x_2^b) &= \begin{cases}1&\text{if } \{x_1^a, x_2^b\} \in
        \mathcal A(X)\\0&\text{otherwise.}\end{cases}
\end{align}
%

Given two genomes $G$ and $H$ and gene similarity measure $\sigma$, two
adjacencies, $\{g_1^a, g_2^b\} \in \mathcal A(G)$ and $\{h_1^a, h_2^b\} \in
\mathcal A(H)$ with $a, b \in \{\textnormal{h, t}\}$ are \emph{conserved} iff
$\sigma(g_1, h_1) > 0$ and $\sigma(g_2, h_2) > 0$.  We subsequently define the
\emph{adjacency score} of any four extremities $g^a, h^b, i^c,j^d$, where
$a,b,c,d \in \{\text{h, t}\}$ and $g, h, i, j \in \Sigma$ as the geometric
mean of their corresponding gene similarities:
\begin{align}\label{eqn:adj_score}
    s(g^a, h^b, i^c, j^d) \equiv \sqrt{\sigma(g, h) \cdot \sigma(i, j)}
\end{align}

\paragraph{Median genome, genes and adjacencies.}
Informally, the family-free median problem asks for a fourth genome $M$ that
maximizes the sum of pairwise adjacency scores to three given extant genomes
$G$, $H$, and $I$. In doing so, the gene content of the requested median $M$
must first be defined: each gene $m \in \mathcal C(M)$ must be unambiguously
associated with a triple of extant genes $(g, h, i)$, $g \in \mathcal C(G)$, $h
\in \mathcal C(H)$, and $i \in \mathcal C(I)$. Moreover, we want to associate to a median gene $m$ a sequence similarity score $(g, h, i)$ relatively to  the three extant genes it is related to. As the sequence of the median gene is obviously not available, we define this score as the
geometric mean of their pairwise similarities:
\begin{equation}\label{eqn:gene_score}
    \sigma(g, m) = \sigma(h, m) = \sigma(i, m) \equiv \sqrt[3]{\sigma(g, h) \cdot \sigma(g,
i) \cdot \sigma(h, i)}
\end{equation}

\vspace*{-5mm}
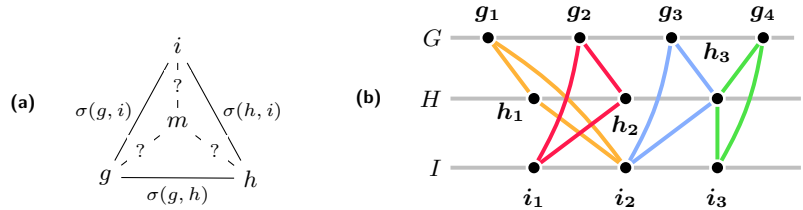
\begin{figure}[h!]
    \begin{center}
    \begin{tikzpicture}[nlabel/.style={font=\scriptsize\sffamily\bfseries\boldmath,anchor=north}]
        \path [line width=1] (0,0) node[anchor=north east] (g) {$g$} --
        (60:1.5) node[anchor=south] (i) {$i$} -- (1.5,0) node[anchor=north
        west] (h) {$h$} -- cycle;

        \node[anchor=north] (m) at (0.75,0.7) {$m$};
        \draw (g) -- node[anchor=north,font=\scriptsize] {$\sigma(g, h)$} (h);
        \draw (g) -- node[anchor=east,font=\scriptsize] {$\sigma(g, i)$}(i);
        \draw (h) -- node[anchor=west,font=\scriptsize] {$\sigma(h, i)$}(i);
        \draw (g) -- node[font=\scriptsize,fill=white] {?} (m);
        \draw (h) -- node[font=\scriptsize,fill=white] {?} (m);
        \draw (i) -- node[font=\scriptsize,fill=white] {?} (m);

        \node[nlabel,left=1.8 of m, anchor=south] {(a)};

    \end{tikzpicture}\hspace{2em}
    \begin{tikzpicture}[node/.style={fill,circle,inner sep=1.5,outer
        sep=1,color=black}, nsmall/.style={fill,circle,inner sep=1,outer
        sep=1,color=black}, adj/.style={line width=1.5,color=gray!50},
        edg/.style={line width=1.5},
        nlabel/.style={font=\scriptsize\sffamily\bfseries\boldmath,anchor=north},
        glabel/.style={font=\footnotesize\sffamily\boldmath,text
        depth=0pt,anchor=south}, node distance=1,auto,every loop/.style={}]
        \node[node] (g1) {};
        \node[node, right=1 of g1] (g2) {};
        \node[node, right=1 of g2] (g3) {};
        \node[node, right=1 of g3] (g4) {};
        \coordinate[right=0.5 of g1] (gi);
        \node[node, below=0.7 of gi] (h1) {};
        \node[node, right=1 of h1] (h2) {};
        \node[node, right=1 of h2] (h3) {};
        \node[node, below=0.7 of h1] (i1) {};
        \node[node, right=1 of i1] (i2) {};
        \node[node, right=1 of i2] (i3) {};

        \coordinate[left=0.39 of g1] (g0);
        \coordinate[left=1 of h1] (h0);
        \coordinate[left=1 of i1] (i0);

        \coordinate[right=0.3 of g4] (gend);
        \coordinate[right=1 of h3] (hend);
        \coordinate[right=1 of i3] (iend);

        \path[adj] (g1) edge (g2);
        \path[adj] (g2) edge (g3);
        \path[adj] (g3) edge (g4);
        \path[adj] (h1) edge (h2);
        \path[adj] (h2) edge (h3);
        \path[adj] (i1) edge (i2);
        \path[adj] (i2) edge (i3);

        \draw[adj] (g0) -- (g1);
        \draw[adj] (h0) -- (h1);
        \draw[adj] (i0) -- (i1);

        \draw[adj] (g4) -- (gend);
        \draw[adj] (h3) -- (hend);
        \draw[adj] (i3) -- (iend);

        \draw[edg,color=a] (g1) -- (h1);
        \draw[edg,color=a] (h1) -- (i2);
        \path[edg,color=a,bend left=10] (g1) edge (i2);

        \draw[edg,color=c] (g2) -- (h2);
        \draw[edg,color=c] (h2) -- (i1);
        \path[edg,color=c,bend left=10] (g2) edge (i1);

        \draw[edg,color=h] (g3) -- (h3);
        \draw[edg,color=h] (h3) -- (i2);
        \path[edg,color=h,bend left=10] (g3) edge (i2);

        \draw[edg,color=l] (g4) -- (h3);
        \draw[edg,color=l] (h3) -- (i3);
        \path[edg,color=l,bend left=10] (g4) edge (i3);


        \node[anchor=east] at (g0) {$G$};
        \node[anchor=east] at (h0) {$H$};
        \node[anchor=east] at (i0) {$I$};

        \node[glabel,above=0.05 of g1]  {$g_1$};
        \node[glabel,above=0.05 of g2]  {$g_2$};
        \node[glabel,above=0.05 of g3]  {$g_3$};
        \node[glabel,above=0.05 of g4]  {$g_4$};

        \node[glabel,below=0.05 of h1, anchor=east]  {$h_1$};
        \node[glabel,below=0.0 of h2]  {$h_2$};
        \node[glabel,above=0.3 of h3]  {$h_3$};

        \node[glabel,below=0.05 of i1] {$i_1$};
        \node[glabel,below=0.05 of i2]  {$i_2$};
        \node[glabel,below=0.05 of i3] {$i_3$};

        \node[nlabel,left=0.8 of h0]{(b)};
    \end{tikzpicture}
    \end{center}
    \vspace*{-8mm}
    \caption{(a) Illustration of the score of a candidate median gene. (b) Gene similarity graph of three genomes $G$, $H$, and $I$. Colored components indicate candidate median genes $m_1 = (g_1, h_1, i_2)$, $m_2 = (g_2, h_2, i_1)$, $m_3 = (g_3, h_3, i_2)$, and $m_4 = (g_4, h_3, i_3)$. Median gene pairs $m_1, m_3$ and $m_3, m_4$ are conflicting.}\label{fig:median_genes}
\end{figure}

In the following we make use of mapping $\pi_G(m) \equiv g$, $\pi_H(m) \equiv
h$, and $\pi_I(m) \equiv i$ to relate gene $m$ with its extant counterparts. Two
candidate median genes or telomeres $m_1$ and $m_2$ are \emph{conflicting} if
$m_1 \neq m_2$ and the intersection between associated gene sets $\{\pi_G(m_1),
\pi_H(m_1), \pi_I(m_1)\}$ and $\{\pi_G( m_2), \pi_H(m_2), \allowbreak
\pi_I(m_2)\}$ is non-empty. A set of candiate median genes or telomeres
$\mathcal C$ is called \emph{conflict-free} if no two of its members $m_1, m_2
\in \mathcal C$ are conflicting. This definition trivially extends to
the notion of a \emph{conflict-free} median.



\begin{problem}[FF-Median]\label{prb:ff-median}
    Given three genomes $G$, $H$, and $I$, and gene similarity measure $\sigma$,
    find a conflict-free median $M$, which maximizes the following formula: 
    \begin{equation}\label{eqn:ff-median-obj}
        \mathcal F_{\Yup}(M) = \sum_{\{m_1^a, m_2^b\} \in \mathcal A(M)} \quad
        \sum_{\substack{X \in \{G, H, I\},\\\{\pi_X(m_1)^a, \pi_X(m_2)^b\} \in
        \mathcal A(X)}} s(m_1^a, \pi_X(m_1)^a, m_2^b, \pi_X(m_2)^b),
    \end{equation}
    \noindent where $a,b \in \{\text{h}, \text{t}\}$ and $s(\cdot)$ is the
    adjacency score as defined by Equation~(\ref{eqn:adj_score}). 
\end{problem}

\begin{remark}\label{rk:score_adj}
The adjacency score for a median adjacency $\{m_1^a, m_2^b\}$ with respect to the corresponding potential extant adjacency  $\{\pi_X(m_1)^a,
\pi_X(m_2)^b\}$, where $\{m_1^a, \allowbreak m_2^b\} \in \mathcal A(M)$ and $X
\in \{G, H, I\}$, can be entirely expressed in terms of pairwise similarities
between genes of extant genomes 
using Equation~(\ref{eqn:gene_score}):
{\small
\begin{alignat*}{1}
    s(m_1^a, \pi_X(m_1)^a, m_2^b, \pi_X(m_2)^b) =  \sqrt[6]{\prod_{\{Y, Z\}
    \subset \{G, H, I\}} \sigma(\pi_Y(m_1), \pi_Z(m_1)) \cdot \sigma(\pi_Y(m_2),
    \pi_Z(m_2))} 
\end{alignat*}
}
\end{remark}

In the following, a median gene $m$ and its extant counterparts $(g, h, i)$ are
treated as equivalent. We denote the set of all \emph{candidate median genes}
by
\begin{equation}
\Sigma_\Yup = \{(g, h, i)~|~g \in \mathcal C(G), h \in \mathcal
C(H), i \in \mathcal C(I): \sigma(g, h) \cdot \sigma(g , i) \cdot
\sigma(h, i) > 0\}\,.
\end{equation}
Each pair of median genes $(g_1, h_1, i_1), (g_2, h_2, i_2) \in \Sigma_\Yup$
and extremities $a, b \in \{\text{h, t}\}$ give rise to a \emph{candidate median
adjacency} $\{(g_1^a, h_1^a, i_1^a), (g_2^b, h_2^b, i_2^b)\}$ if $(g_1^a,
h_1^a, i_1^a) \neq (g_2^b, h_2^b, i_2^b)$, and $(g_1^a, h_1^a, i_1^a)$ and
$(g_2^b, h_2^b, i_2^b)$ are non-conflicting. We denote the set of all candidate
median adjacencies and the set of all \emph{conserved} (\textit{i.e.} present in at least one extant genome) candidate median
adjacencies by $\mathcal A_\Yup = \{\{m_1^a, m_2^a\}~|~m_1,m_2 \in
\Sigma_\Yup,~a,b \in \{\text{h}, \text{t}\}\}$ and $\mathcal A_\Yup^C =
\{~\{m_1^a, m_2^b\} \in \mathcal A_\Yup~|~\allowbreak\sum_{X \in \{G, H, I\}}
\allowbreak \mathbb I_X(\pi_X(m_1)^a, \pi_X(m_2)^b) \geq 1\}$, respectively.

\begin{remark}\label{rk:cliques}
A median gene can only belong to a median adjacency with non-zero adjacency score if all pairwise similarities of its corresponding extant genes $g, h, i$ are non-zero. Thus, the search for median genes can be limited to  $3$-cliques (triangles) in the tripartite similarity graph.
\end{remark}

\begin{remark}\label{rk:conserved_adjs}
The right-hand side of the above formula for the weight of an adjacency is independent of genome $X$. From Equation~(\ref{eqn:ff-median-obj}), an
adjacency in median $M$ has only an impact in a solution to problem FF-Median
if it participates in a gene adjacency in at least one extant
genome. So including in a median genome median genes that do not belong to a candidate median adjacency in $\mathcal A_\Yup^C$  do not increase the objective function. 
\end{remark}

\paragraph{Related problems.}
The FF-median problem relates to previously studied gene order evolution problems. It is a generalization of the tractable mixed multichromosomal median problem introduced in~\cite{tannier2009zs}, that can indeed be defined as an FF-median problem with a similarity graph composed of disjoint $3$-cliques and edges having all the same weight. 
The FF-median problem also bears similarity with methods aimed at detecting
groups of orthologous genes based on gene order evolution, especially the
MultiMSOAR~\cite{Shi:2011cch} algorithm, although other method integrate synteny and sequence conservation for inferring orhogroups, see~\cite{Dewey:2011bp}. Our approach differs first and foremost in its family-free principle (all other methods require a prior gene family assignment).  Compared to MultiMSOAR, the only other method that can handle  more than two genomes with an optimization criterion that considers gene order evolution, both MultiMSOAR (for three genomes) and FF-median aim at computing a maximum weight tripartite matching. However we differ fundamentally from MultiMSOAR by the full integration of sequence and synteny conservation into the objective function, while MultiMSOAR proceeds first by computing pairwise orthology assignments to define a multipartite graph.

\section{Algorithmic and complexity results}~\label{sec:complexity}

\vspace*{-5mm}
We now describe our theoretical results: a NP-hardness proof, an exact Integer Linear Program (ILP), and an algorithm to detect local optimal structures.


\begin{theorem}\label{thr:maxsnp_hardness}
    Problem FF-Median is MAX~SNP-hard.
\end{theorem}

We describe the full hardness proof in Appendix \ref{app:hardness_proof}. It is
based on a reduction from the Maximum Independent Set for Graphs of Bounded
Degree $3$. 

\paragraph{An exact ILP algorithm to problem FF-Median.}\label{sec:algorithms}
We now present program
\texttt{FF-Median}, described by Algorithm~\ref{alg:ff-median}, that exploits the specific properties of problem
 FF-Median to design an ILP using $\mathcal O(n^5)$ variables and statements. 
Program \texttt{FF-Median} makes use of two types of binary variables $\mathbf
a$ and $\mathbf b$ as declared in domain specifications (\texttt{D.01}) and
(\texttt{D.02}), that defines the set of median genes $\Sigma_\lambda$ and of candidate conserved median adjacencies ${\mathcal A_\Yup^C}$ (Remark~\ref{rk:conserved_adjs}). The former variable type indicates the presence or absence of
candidate genes in an optimal median $M$. The latter, variable type $\mathbf
b$, specifies if an adjacency between two gene extremities or telomeres is
established in $M$. 
Constraint (\texttt{C.01}) ensures that $M$ is conflict-free, by demanding that
each extant gene (or telomere) can be associated with at most one median gene
(or telomere).  Further, constraint (\texttt{C.02}) dictates that a median
adjacency can only be established between genes that both are part of the
median. Lastly, constraint (\texttt{C.03}) guarantees that each gene extremity
and telomere of the median participates in at most one adjacency. 

\begin{algorithm}[!h]
	\caption{
        Program \texttt{FF-Median} for three genomes $(G,H,I)$}
        \label{alg:ff-median}	
	\begin{minipage}{\columnwidth}{ 
       \noindent\textbf{Objective:}       
        \texttt{Maximize }\smallskip\\ {\scriptsize
        $\displaystyle\sum_{ {\scriptsize\substack{(g_1, h_1, i_1), \\(g_2, h_2,
        i_2) \in {\mathcal A_\Yup^C},\\a, b \in \{\text{h, t}\}}}}
        \mathbf b(g_1^a,g_2^b, h_1^a,h_2^b, i_1^a, i_2^b)~
        \displaystyle\sqrt[6]{\sigma(g_1, h_1, i_1) \sigma(g_2, h_2, i_2)}~(\mathbb I_G(g_1^a, g_2^b)~+~\mathbb I_H(h_1^a, h_2^b)~+~\mathbb
        I_I(i_1^a, i_2^b))$}

		\smallskip
		
        \noindent\textbf{Constraints:}
		\smallskip

        \noindent\;
        (\texttt{C.01})\quad $\forall~g' \in \mathcal C(G)$:
        $\displaystyle\sum_{\substack{(g, h, i) \in \Sigma_\Yup,\ g =
        g'}} \mathbf a(g, h, i) \leq 1$\smallskip

        \noindent\;
        \phantom{(\texttt{C.01})\quad} $\forall~h' \in \mathcal C(H)$:
        $\displaystyle\sum_{\substack{(g, h, i) \in \Sigma_\Yup,\ h =
        h'}} \mathbf a(g, h, i) \leq 1$\smallskip

        \noindent\;
        \phantom{(\texttt{C.01})\quad} $\forall~i' \in \mathcal C(I)$:
        $\displaystyle\sum_{\substack{(g, h, i) \in \Sigma_\Yup,\ i =
        i'}} \mathbf a(g, h, i) \leq 1$\bigskip

        \noindent\;
        (\texttt{C.02})\quad $\forall~(g_1, h_1, i_1), (g_2, h_2, i_2) \in
        \Sigma_\Yup$ and $\forall~a, b \in \{\text{h, t}\}$:
        \smallskip
 
        \noindent\;
        \phantom{(\texttt{C.02})}\quad\quad\quad$2\cdot\mathbf b(g_1^a, g_2^b, h_1^a,
        h_2^b, i_1^a, i_2^b) \leq \mathbf a(g_1, h_1, i_1) + \mathbf a(g_2,
        h_2, i_2)$\bigskip
        
        \noindent\;
        (\texttt{C.03})\quad $\forall~(g_1, h_1, i_1) \in \Sigma_\Yup$
        and $\forall~a \in \{\text{h, t}\}$: \smallskip
        
        \noindent\; \phantom{(\texttt{C.03})\quad\quad\quad}
        $\displaystyle\sum_{\substack{(g_2, h_2, i_2) \in \Sigma_\Yup,\ b
        \in \{\text{h, t}\} }} \mathbf b(g_1^a, g_2^b, h_1^a, h_2^b, i_1^a,
        i_2^b) \leq 1$\smallskip

		\smallskip
		\noindent\;\textbf{Domains:}
		\smallskip
		
        \noindent\; 
        (\texttt{D.01})\quad $\forall~(g, h, i) \in \Sigma_\Yup$:
        ~$\mathbf a(g, h, i) \in \{0, 1\}$\smallskip

        \noindent\; 
        (\texttt{D.02})\quad $\forall~(g_1, h_1, i_1), (g_2, h_2, i_2) \in
        {\mathcal A_\Yup^C}$ and $\forall~a, b \in \{\text{h, t}\}$: 
        \smallskip

        \noindent\;
        \phantom{(\texttt{D.02})\quad}\quad\quad $\mathbf b(g_1^a, g_2^b, h_1^a,
        h_2^b, i_1^a, i_2^b) \in \{0, 1\}$\bigskip
        }\end{minipage}
\end{algorithm}

\begin{property}
	The size (i.e. number of variables and statements) of any ILP returned by program \texttt{FF-Median} is limited by $\mathcal O(n^5)$ where $n=\max(|\mathcal C(G)|,|\mathcal C(H)|,|\mathcal C(I)|)$.
\end{property}

\begin{remark}\label{rk:output}
The output of the algorithm \texttt{FF-Median} is a set of adjacencies between median genes that  define a set of linear and/or circular orders, called CARs (Contiguous Ancestral Regions),  where linear segments are not capped by telomeres. So formally the computed median might not be a valid genome. However, as adding  adjacencies that do not belong to $\mathcal A_\Yup^C$ do not modify the score of a given median, a set of median adjacencies can always be completed into a valid genome by such adjacencies that join the linear segments together and add telomeres. These extra adjacencies would not be supported by any extant genome and thus can be considered as dubious, and in our implementation, we only return the median adjacencies computed by the ILP, \textit{i.e.} a subset of $\mathcal A_\Yup^C$.
\end{remark}

\begin{remark}\label{rk:non_clique}
Following Remark~\ref{rk:cliques}, preprocessing the input extant genomes requires to handle the extant genes that do not belong to at least one $3$-clique in the similarity graph. Such genes can not be part of any median. So one could decide to leave them in the input, and the ILP can handle them and ensures they are never part of the output solution. However, discarding them from the extant genomes can help recover adjacencies that have been disrupted by the insertion of a mobile element for example, so in our implementation we follow this approach. 
\end{remark}

As discussed at the end of Section~\ref{sec:ffmedian}, the FF-median problem is a generalization of the mixed multichromosomal breakpoint median~\cite{tannier2009zs}. 
However, it was shown in~\cite{tannier2009zs} that this breakpoint median problem can be solved in polynomial time by a Maximum-Weight Matching (MWM) algorithm. This motivates the results presented in the next paragraph that use a MWM algorithm to identify optimal median substructures by focusing on conflict-free sets of median genes.

\paragraph{Finding local optimal segments.}

Tannier~\etal~\cite{tannier2009zs} solve the mixed multichromosomal breakpoint median problem by transforming it into an MWM problem, that we outline now. A graph is defined in which each extremity of a candidate median gene and each telomere gives rise to a vertex. Any two vertices are connected by an edge, weighted according to the number of observed adjacencies between the two gene extremities in extant genomes. Edges corresponding to adjacencies between a gene extremity and telomeres are weighted only by half as much. An MWM in this graph induces a set of adjacencies that defines an optimal median.

We first describe how this approach applies to our problem. We define a graph $\Gamma(G, H, I, \sigma)$ constructed from an FF-Median instance $(G, H, I, \sigma)$ that is similar to that of Tannier~\etal, only deviating by defining vertices as candidate median genes and weighting an edge between two candidate median gene extremities (or telomeres) $m_1^a, m_2^b$, $a, b \in \{\textnormal{h, t}\}$, by

\begin{align}\label{eq:weights}
\begin{split}
w(\{m_1^a, m_2^b\}) &= \sqrt[6]{\prod_{\{X, Y\} \subset \{G, H,
I\}} \sigma(\pi_X(m_1), \pi_Y(m_1))\sigma(\pi_X(m_2), \pi_Y(m_2))}\\
&  \cdot \displaystyle\sum_{X \in \{G, H, I\}} \mathbb I_X(\pi_X(m_1)^a,
\pi_X(m_2)^b).
\end{split}
\end{align}

We make first the following observation, where a conflict-free matching is a matching that does not contain two conflicting vertices (candidate median genes):

\begin{observation}\label{obs:MWM}
    Any conflict-free matching in graph $\Gamma(G, H, I, \sigma)$ of maximum weight  defines an optimal median. 
\end{observation}

We show now that we can define notions of sub-instances -- of a full FF-median instance -- that contains no  internal conflicts, for which applying the MWM can allow to detect if the set of median genes defining the sub-instance is part of at least one optimal FF-median. 
Let $\mathcal S$ be a set of candidate median genes. An \textit{internal conflict} is a conflict between two genes from $\mathcal S$; an \textit{external conflict} is a conflict between a gene from $\mathcal S$ and a candidate median gene not in $\mathcal S$. We say that $\mathcal S$ is \textit{contiguous} in extant genome $X$ if the set $\pi_X({\mathcal S})$ forms a unique, contiguous, segment in $X$. We say that $\mathcal S$ is an \textit{internal-conflict free segment} (IC-free segment) if it contains no internal conflict and is contiguous in all three extant genomes; this can be seen as the family-free equivalent of the notion of \textit{common interval in permutations}~\cite{Bergeron:2008cch}. An IC-free segment is \textit{framed} if the extremities of the extant segments belong to the same two median genes, with conserved relative orientations (the equivalent of a \textit{conserved interval}). An IC-free segment is a \textit{run} if the order of the extant genes is conserved in all three extant genomes, up to a full reversal of the segment.

Intuitively, one can find an optimal solution to the sub-instance defined by an IC-free segment, but it might not be part of an optimal median for the whole instance due to side effects of the rest of the instance. So we need to adapt the graph to which we apply an MWM algorithm to account for such side effects. To do so, we define the
\textit{potential} of a candidate median gene $m$  as 
\[\Delta(m) = \max_{\{\{m_1^a,m^b\},\{m^a, m_2^b\}\} \in \mathcal{A}_\Yup}
\; \big( w(\{m_1^a, m^b\}) + w(\{m^a, m_2^b\})\big).\]

We then extend graph $\Gamma =: (V, E)$ to graph $\Gamma' := (V, E')$ by adding
edges between the extremities of each candidate median gene of an IC-free
segment $\mathcal S$, i.e.~$E'= E \cup \{ \{m^{\textnormal{h}},
m^{\textnormal{t}}\}~|~m \in \mathcal S\}$. In the following we refer to these
edges as \emph{conflict edges}. Let $C(m)$ be the set of candidate median genes
that are involved in an (external) conflict with a given candidate median gene
$m$ of $\mathcal S$, then the conflict edge $\{m^\textnormal{h},
m^\textnormal{t}\} \in E'$ is weighted by the maximum potential of a
non-conflicting subset of $C(m)$, 
\[ w'(\{m^\textnormal{h}, m^\textnormal{t}\}) = \max(\{\sum_{m'\in
C'}\Delta(m')~|~C' \subseteq C(m) :~
C' \text{ is conflict-free}\})\,.\]
A conflict-free matching in $\Gamma'$ is a matching that does not contain a
conflict edge. 

\begin{lemma}\label{lem:ICF-SEG} 
Given an internal conflict-free segment $\mathcal S$, any maximum weight
matching in graph $\Gamma'(S)$ that is conflict-free defines a set of median
genes and adjacencies that belong to at least one optimal FF-median of the whole
instance.
\end{lemma}

A proof is presented in Appendix~\ref{app:eg}. Lemma~\ref{lem:ICF-SEG} leads to a
procedure (Algorithm~\ref{alg:cf_seg}) that
iteratively identifies and tests IC-free segments in the FF-Median instance.
For each identified IC-free segment $S$ an adjacency graph
$\Gamma'(S)$ is constructed and a maximum weight matching is computed (lines
2-3). If the resulting matching is conflict free (line 4), adjacencies of
IC-free segment $S$ are reported and $S$ is removed from an FF-Median instance
by masking its internal adjacencies and removing all candidate median genes (and
consequently their associated candidate median adjacencies) corresponding to
external conflicts (lines 5-6). It then follows immediately from Lemma~\ref{lem:ICF-SEG} that the set median genes returned by Algorithm~\ref{alg:cf_seg} belongs to at least one optimal solution to the FF-median problem.

\begin{algorithm}[h!]
    \caption{Algorithm \texttt{ICF-SEG}}
    \label{alg:cf_seg}
{\footnotesize
\begin{algorithmic}[1]
    \REQUIRE FF-Median instance $(G, H, I, \sigma)$
    \ENSURE Set of adjacencies $\textsc{Adj}_{M}$ that is part of a median $M$
    of $(G, H, I, \sigma)$. 
    \vspace{0.5em}
    \WHILE{there exists an unobserved IC-free conserved segment $S$ in $(G, H, I, \sigma)$} 
        \STATE Construct adjacency graph $\Gamma'(S)$ of $S$
        \STATE Find maximum weight matching $\mathcal M \subseteq E(\Gamma'(S))$
        \IF {$A(S) = \mathcal M$}
        	\STATE Add $A(S)$ to $\textsc{Adj}_{\mathcal M}$
            \STATE Remove $S$ including external conflicts from $(G, H, I, \sigma)$
        \ENDIF
    \ENDWHILE
\end{algorithmic}
    }
\end{algorithm}

\section{Experimental results and discussion}\label{sec:results}

Our algorithms have been implemented in Python and require
CPLEX\footnote{\url{http://www.ibm.com/software/integration/optimization/cplex-optimizer/}};
they are freely available as part of the family-free genome comparison tool
\texttt{FFGC} downloadable at
\url{http://bibiserv.cebitec.uni-bielefeld.de/ffgc}.

In subsequent analyses, gene similarities are based on local alignment hits
identified with BLASTP~\cite{Altschul:1990dw} on protein sequences using an
e-value threshold of $10^{-5}$. In  gene similarity graphs, we
discard spurious edges by applying a \emph{stringency filter} proposed by
Lechner~\etal~\cite{Lechner:2011jk} that utilizes a local threshold parameter
$f \in [0, 1]$ and BLAST bitscores: a BLAST hit from
a gene $g$ to $h$ is only retained if it is has a higher or equal score than
$f$ times the best BLAST hit from $h$ to any gene $g'$ that is member of the
same genome as $g$. In all our experiments, we set $f$ to $0.5$. Edge weights
of the gene similarity graph are then calculated according to the
\emph{relative reciprocal BLAST score} (RRBS)~\cite{Pesquita:2008jw}. Finally we 
applied Algorithm ICF-SEG with conserved segments defined as runs. 

For solving the FF-Median problem, we granted CPLEX two CPU cores, 4~GB memory, and a time limit of 3 hours per dataset. 

In our experiments, we compare ourselves against the orthology prediction tool
MultiMSOAR~\cite{Shi:2011cch}. This tool requires precomputed gene families,
which we constructed by following the workflow described in \cite{Shi:2011cch}.

\paragraph{Evaluation on simulated data.}

We first evaluate our algorithms on simulated data sets obtained by
ALF~\cite{Dalquen:2012dx}.  The ALF simulator covers many aspects of genome
evolution from point mutations to global modifications. The latter includes two
types of genome rearrangements, as well as various options to customize the
process of gene family evolution.  In our simulations, we mainly use standard
parameters suggested by the authors of ALF and we focus on three parameters
that primarily influence the outcome of gene family-free genome analysis: (i)
the rate of sequence evolution, (ii) the rate of genome rearrangements, and
(iii) the rate of gene duplications and losses. We keep all three rates
constant, only varying the evolutionary distance between the generated extant
genomes.  We confine our simulations to protein coding sequences.  A
comprehensive list of parameter settings used in our simulations is shown in
Table~\ref{tab:alf_params} in Appendix~\ref{app:simulations}.  As root genome
in the simulations, we used the genomic sequence of an \emph{E.~coli}
\mbox{K-12} strain\footnote{Accession no: \texttt{NC\_000913.2}} which comprises $4,320$
protein coding genes.  We then generated  $7\times 10$ data sets with
increasing evolutionary distance ranging from 10 to 130 \emph{percent accepted
mutations} (PAM). Details about the generated data sets are shown in Table
\ref{tab:median_data_info} in Appendix~\ref{app:simulations}.
Figure~\ref{fig:eval}~(a) shows the outcome of our analysis with respect to
precision and recall\footnote{precision: \#true positives/(\#true positives +
\#false positives), recall: \#true positives/(\#true positives + \#false
negatives)} of inferring positional orthologs. In all simulations, FF-Median
generated no or very few false positives, leading to perfect or near-perfect
precision score, consistently outperforming MultiMSOAR. However, since the
objective of FF-Median only takes median genes into account that are conserved
by synteny, the increase in mutational changes over evolutionary time causes a
growing loss of syntenic context which results in a lower recall. Therefore,
MultiMSOAR retains a better recall for larger evolutionary distances, while FF-Median provides better results for more closely related genomes. 

\begin{figure}[tb]
    \begin{center}
    \includegraphics[width=0.45\columnwidth]{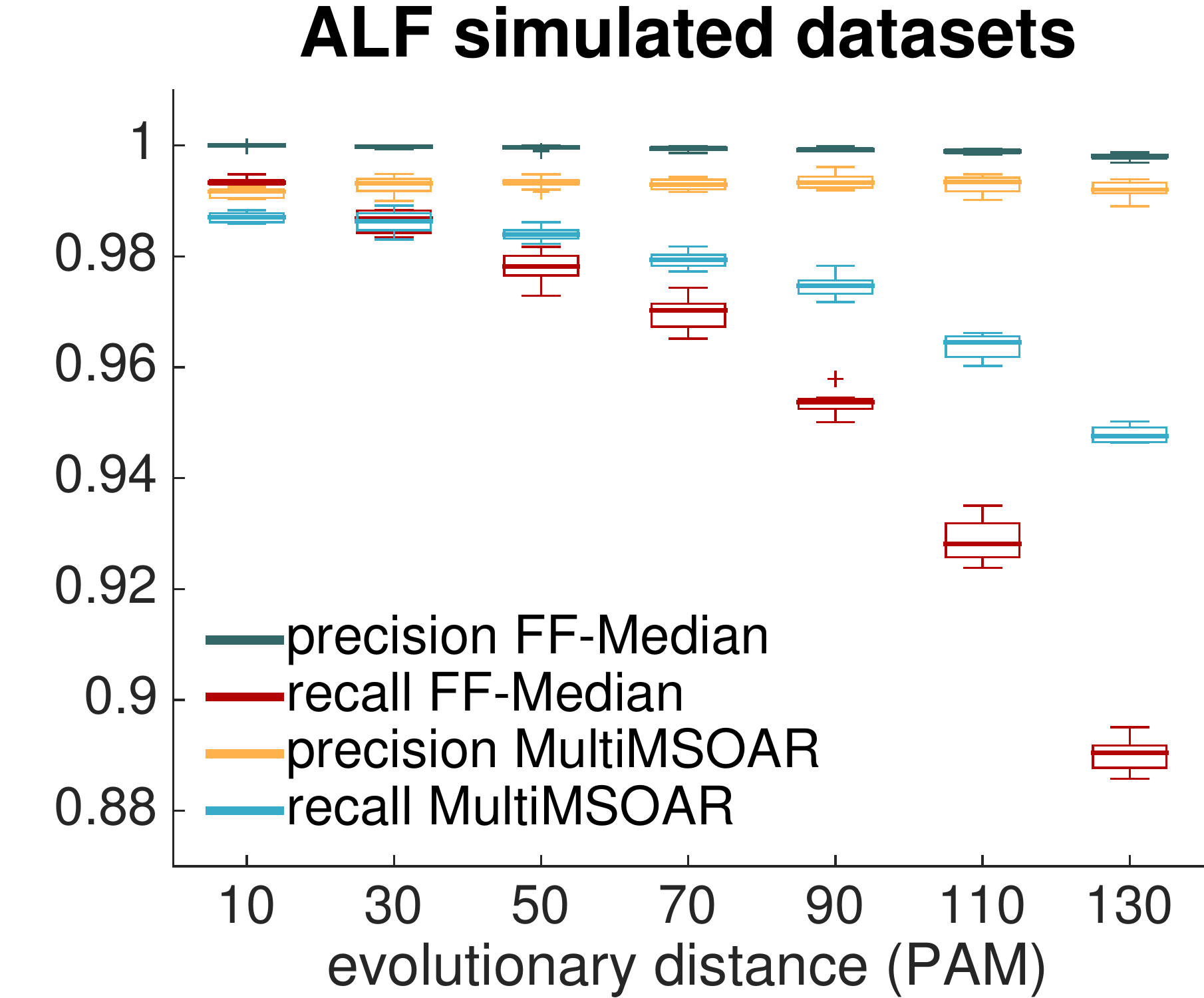}
    \includegraphics[width=0.45\columnwidth]{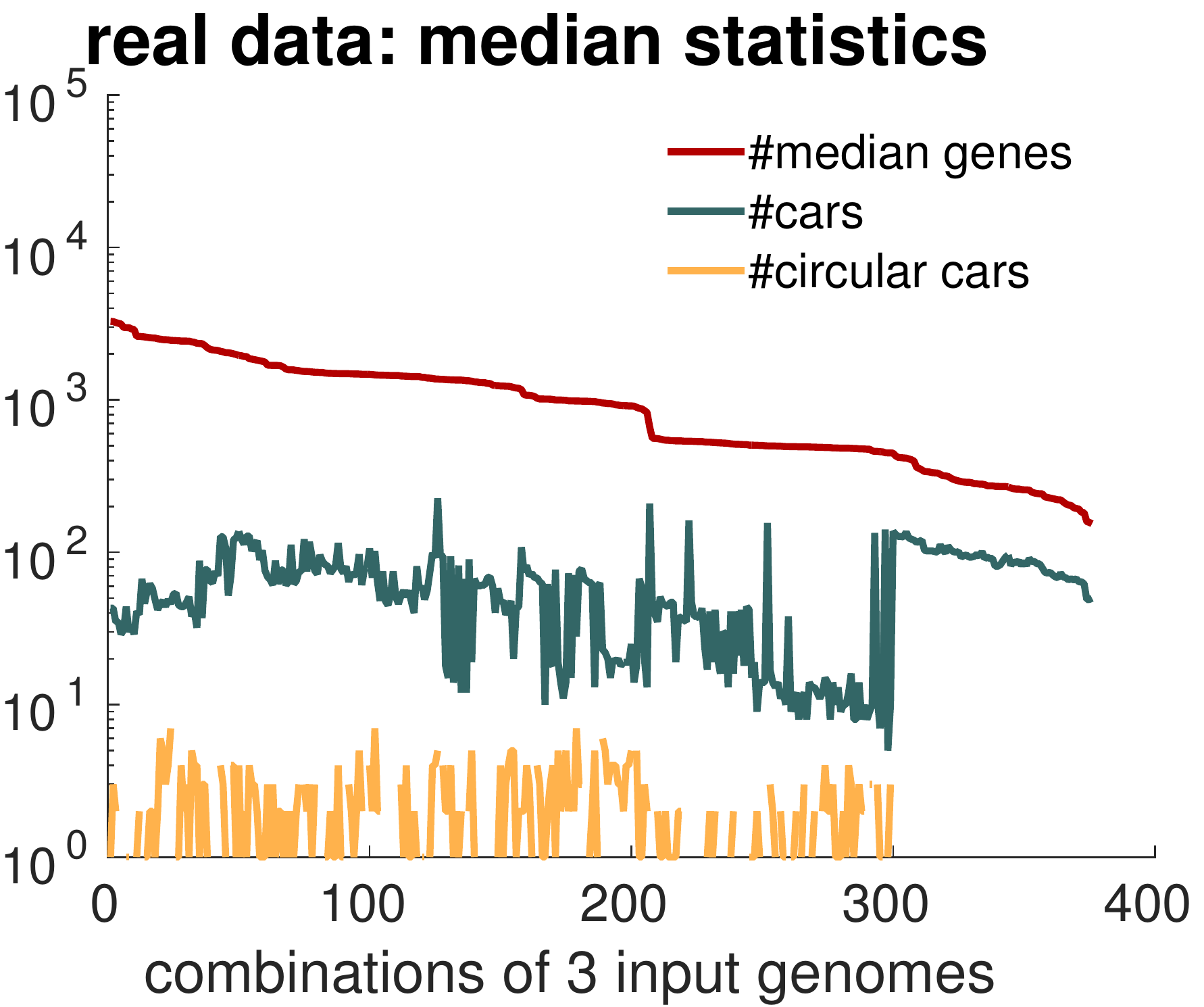}

    {\scriptsize \hfill(a)\hfill(b)\hfill\phantom{.}}
    \medskip

    \includegraphics[width=0.45\columnwidth]{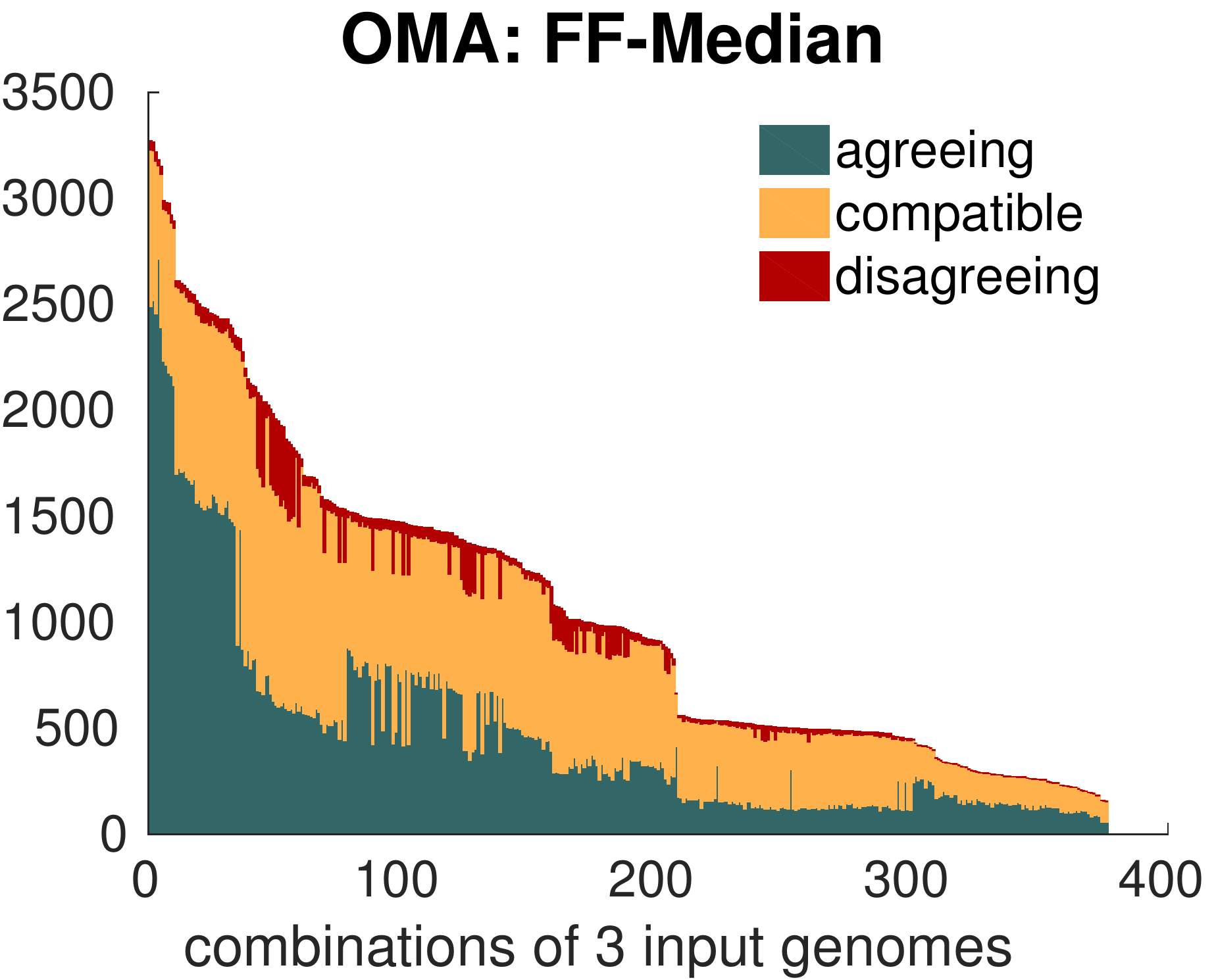}
    \includegraphics[width=0.45\columnwidth]{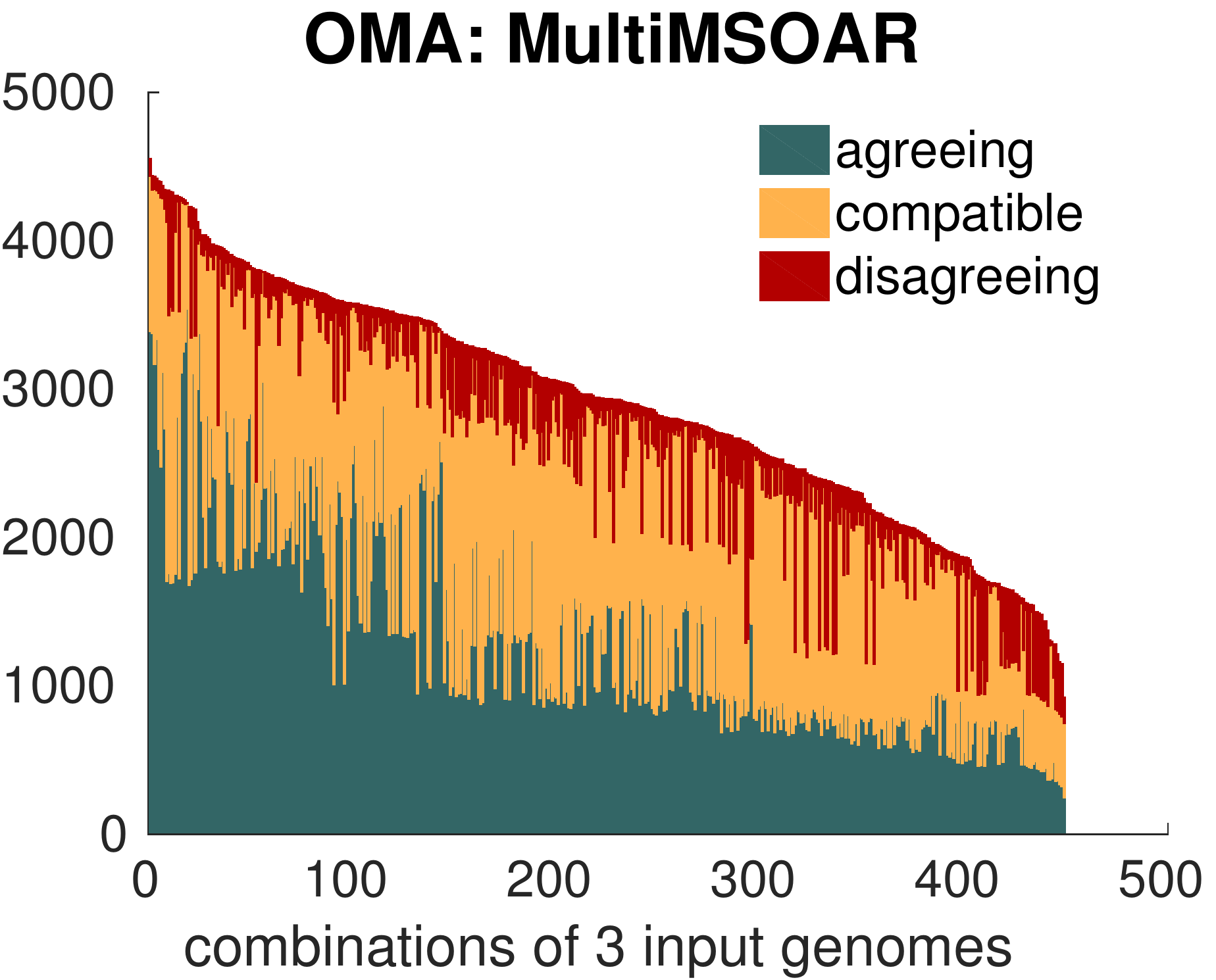}
    
    {\scriptsize \hfill(c)\hfill(d)\hfill\phantom{.}}

    \end{center}
    \caption{Top: (a) Precision and recall of FF-Median and MultiMSOAR
    in simulations; (b) statistical assessment of CARs and median genes on real
    datasets.  Bottom: agreement, compatiblity and disagreement of
    positional orthologs inferred by (c) FF-Median and (d) MultiMSOAR with
    OMA database.}\label{fig:eval}
\end{figure}

\paragraph{Evaluation on real data.}

We study 15 $\gamma$-proteobacterial genomes that span a large taxonomic spectrum and are contained in the OMA database~\cite{Altenhoff:2015}. A complete list of species names is given in Appendix~\ref{app:real_genomes}. We obtained the genomic sequences from the NCBI database and constructed for each combination of three genomes a gene similarity graph following the same procedure as in the simulated dataset.
In $9$ out of the $455$ combinations of genomes the time limit prohibited CPLEX
from finding an optimal solution. However, in those cases CPLEX was
still able to find integer feasible suboptimal solutions. Figure~\ref{fig:eval}~(b) displays
statistics of the real dataset. The number of candidate median genes and adjacencies ranges from $442$ to $18,043$ and $3,164$ to $2,261,716$, respectively, giving rise to up to $3,227$ median genes that are distributed on $5$ to $91$ CARs per median. Some  CARs are circular, indicating dubious conformations mostly arising from tandem duplications, but the number of such cases were low (mean: $2.78$, max: $13$). 

We observed that the gene families in the OMA database are clustered 
tightly and therefore missing many true orthologies in the considered triples of genomes. As a result, many of the
orthologous groups inferred by FF-Median and MultiMSOAR fall into more than one
gene family inferred by OMA. We therefore evaluate our results by
classifying the inferred orthologous groups into three categories: 
An orthologous group \emph{agrees} with OMA if its three genes are in the same OMA group. It
\emph{disagrees} with OMA if extant genes $x$ and $y$ (of genomes $X$ and $Y$
respectively) are in different OMA groups but the OMA group of $x$ contains
another gene from genome $Y$. It is \emph{compatible} with OMA if it neither agrees nor
disagrees with OMA. We measure the number of median genes as well as the
number orthologous groups of MultiMSOAR in each of the three categories.
Figure~\ref{fig:eval}~(c) and (d) show the outcome this analysis. MultiMSOAR is
generally able to find more orthology relations in the dataset. This comes at
no surprise, as it is clear from the objective of problem FF-Median and from
the results of the simulated datasets that our method does not retain candidate
median genes which have lost their syntenic context, which happens in triples of highly divergent genomes. The number of disagreeing  orthologous groups that disagree with OMA is comparably low for both FF-Median (mean: $35.16$, var: $348$) and MultiMSOAR (mean: $48.61$, var: $348$).


We then performed another analysis to assess the \emph{robustness} of the positional orthology predictions. To this end, we look at orthologous groups across multiple datasets that share two extant genomes, but vary in the third. Given two genes, $x$ of genome $X$ and $y$ of genome $Y$, an orthologous group that contains $x$ and $y$ is called \emph{robust} if $x$ and $y$ occur in the same orthologous group, whatever the third extant genome is. We computed the percentage of robust orthologous groups for all gene pairs of genomes \emph{E. coli K-12 MG 1655} and \emph{S. enterica subsp. enterica serovar Typhimurium str. 14028s} in our dataset. The results indicate that orthologous groups inferred by FF-Median are slightly more robust ($95.61\%$) than robust those by MultiMSOAR ($91.77\%$). This is likely due to the strict constraint of defining median adjacencies only from genes that participate in at least one observed adjacency (Remark~\ref{rk:output}). 

Overall, we can observe that FF-Median performed better than MultiMSOAR only for triples of closely related genomes -- which is consistent with our observation on simulated data -- while being slightly more robust in general. This suggests FF-Median is an interesting alternative to identify higher confidence positional orthologs, at the expense of a higher recall rate.

\vspace*{-5mm}
\paragraph{Future work.}
We first aim to investigate alternative methods to reduce the
computational load of Program FF-Median by identifying further strictly
sub-optimal and optimal substructures, which might require to understand better the impact of internal conflicts within substructures defined by intervals in the extant genomes. Without the need to modify drastically either the FF-median problem definition or the ILP, one can think about more complex weighting schemes for adjacencies that could account for known divergence time between genomes or relaxed notion of adjacencies that would address the high recal rate we observe in FF-Median. Within that regard, it would probably be interesting to combine this with the use of common intervals instead of runs to define conflict-free sub-instances.
Finally, ideal family-free analysis should take into account the effects of gene family evolution. However, the presented family-free median model can only resolve certain cases
of gene duplication. It is generally susceptible to gene losses that occurred
along the evolutionary paths between the three extant genomes and their common ancestor. The
definition of a family-free median model that tolerates events of gene family
evolution at a reasonable computational cost  is likely an interesting research avenue.


%

\bibliographystyle{plain}
\bibliography{references.bib}

\newpage
\appendix
\FloatBarrier

\section{Hardness proof}\label{app:hardness_proof}

\subsection{Reduction}

The \emph{maximum independent set problem for graphs bounded by node degree
$3$}, denoted as MAX~IS-3 is MAX~SNP-hard~\cite{Papadimitriou:1991fj}. The
corresponding decision problem can be informally stated as follows: Given a
graph $\Lambda$ bounded by degree 3 and some number $l \geq 1$, does there
exists a set of vertices $V' \subseteq V$ of size $|V'| = l$ whose induced
subgraph is unconnected? In the following, we present a transformation scheme
\textbf{R} to phrase $\Lambda$ as FF-median instance $\mathbf{R}(\Lambda) = (G,
H, I, \sigma)$ such that the value $\mathcal F_\Yup(M)$ of a median $M$ of
$\mathbf R(\Lambda)$ is limited by $\mathcal F_\Yup(M) \leq 2 \cdot l + 3$. In
doing so, we associate vertices of $V$ with genes of extant genomes $G, H$ and
$I$. In order to keep track of associated genes, we denote by function
$\xi(x)$ the list of vertices associated with gene $x$.  We further introduce
two types of unassociated genes, ``$\emptyset$'' and ``$\ast$'', whose members
are identified by subscript notation.

\bigskip

\noindent Transformation \textbf{R}:
\begin{enumerate}
    \item Construct genome $G$ such that for each vertex $v \in V$ there exists
        two associated genes $g_v, \bar g_v \in \mathcal C(G)$, i.e.~$\xi(g_v)
        = \xi(\bar g_v) = v$. Further, let each gene pair $g_v, \bar g_v$ form a
        circular chromosome, giving rise to adjacency set $\mathcal A(G) = \{
        \{\bar g_v^\text{h}, g_v^\text{t}\}, \allowbreak \{\bar g_v^\text{h},
        g_v^\text{t}\}~|~v \in V,~g_v, \bar g_v \in \mathcal C(G)\}$. 
    \item For each edge $(u, v) \in E$ construct a circular chromosome
        $\mathcal X_{uv}$ hosting two genes $x_{uv}, x_\emptyset \in \mathcal
        C(\mathcal X_{uv})$, with gene $x_{uv}$ being associated with both
        vertices $u$ and $v$ and gene $x_\emptyset$ being unassociated.
        Further, let both genes form a circular chromosome, giving rise to
        adjacency set $\mathcal A(\mathcal X_{uv}) = \{ \{x_{uv}^\text{h},
        x_\emptyset^\text{t}\}, \{x_\emptyset^\text{h}, x_{uv}^\text{t} \} \}$.
    \item Assign each chromosome constructed in the previous step either to
        genome $H$ or to genome $I$ such that each vertex $v \in V$ is
        associated with at most two genes per genome. 
    \item Complete genomes $H$ and $I$ with additional circular chromosomes
        $\mathcal X_v$ where $\mathcal C(\mathcal X_v) = \{x_v, x_\emptyset\}$
        and $\mathcal A(\mathcal X_v) = \{ \{x_v^\text{h}, x_\emptyset^\text{t}
        \}, \{x_\emptyset^\text{h}, x_v^\text{t} \} \}$ such that each vertex
        in $V$ is associated with exactly two genes per genome.  
    \item For each vertex $v \in V$, let $g, \bar g \in \mathcal C(G)$, $h,
        \bar h \in \mathcal C(H)$, and $i, \bar i \in \mathcal C(I)$ be the
        pairs of genes associated with $v$, i.e.~$\xi(g) = \xi(\bar g) = \xi(h)
        \cap \xi(i)  = \xi(\bar h) \cap \xi(\bar i) = v$.  Assign gene
        similarities $\sigma(g, h) = \sigma(g, i) =  \sigma(h, i) = 1$ and
        $\sigma(\bar g, \bar h)= \sigma(\bar g, \bar i) = \sigma(\bar h, \bar
        i) = 1$.
    \item Add a copy of circular chromosome $\mathcal X_\ast$ to each genome
        $G, H$, and $I$, where $\mathcal C(\mathcal X_\ast) = \{x_\ast, \bar
        x_\ast\}$ and $\mathcal A(\mathcal X_\ast) = \{\{x_\ast^\text{h}, \bar
        x_\ast^\text{t} \}, \{\bar x_\ast^\text{h}, x_\ast^\text{t} \} \}$. Let
        $g_\ast, \bar g_\ast \in \mathcal C(G), h_\ast, \bar h_\ast \in \mathcal
        C(H)$, and $i_\ast, \bar i_\ast \in \mathcal C(I)$, set the gene
        similarity score between all pairs of genes in $\{g_\ast, h_\ast,
        i_\ast\}$ and $\{\bar g_\ast, \bar h_\ast, \bar i_\ast \}$
        respectively, to $1$.  Lastly, set the gene similarity score of all
        pairs of unassociated genes of type ``$\emptyset$'' including genes
        $g_\ast, \bar g_\ast$ to $\frac{1}{4}$.  
\end{enumerate}

\noindent Except for step 3, none of the instructions of transformation scheme
$\mathbf R$ are computationally challenging. Note that in step 3 the demanded
partitioning of chromosomes into genomes $H$ and $I$ is always possible as
consequence of Vizing's Theorem~\cite{Vizing:1964}, by which every graph with
maximum node degree $d$ is edge-colorable using at most $d$ or $d+1$ colors.
Hence, using colors $\chi_1, \chi_2, \chi_3, \chi_4$ with $\chi_1 = \chi_2
\equiv I$, $\chi_3 = \chi_4 \equiv H$ and Misra and Gries'
algorithm~\cite{Misra:1992gb}, edges of graph $\Lambda = (E, V)$ can be
partitioned into two groups in $\mathcal O(|E||V|)$ time implying an assignment
to genomes $H$ and $I$. 

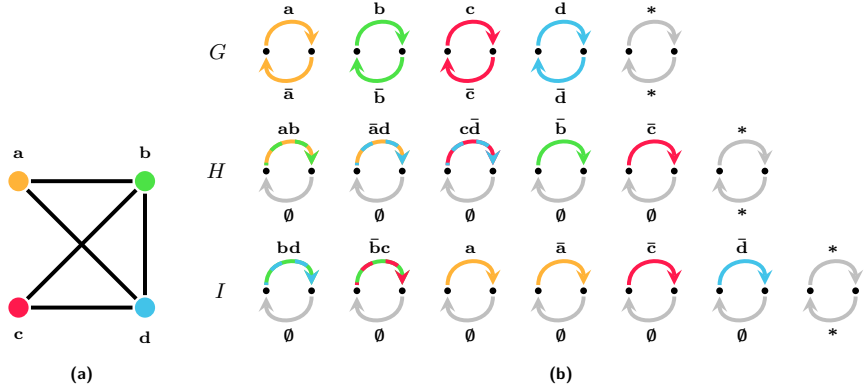
\begin{figure}[tb]
\begin{center}
    \begin{tikzpicture}[scale=0.9, every node/.append style={transform shape},
        node/.style={fill,circle,inner sep=3,outer sep=1,color=black},
        nsmall/.style={fill,circle,inner sep=1,outer sep=1,color=black},
        adj/.style={line width=1.5,color=gray!50}, edg/.style={line width=1.5},
        label/.style={font=\scriptsize\sffamily\bfseries,text
        depth=0pt,anchor=south}, node distance=1,auto,every loop/.style={}]
        \node[node, color=a] (yellow) {};
        \node[node, color=l, right=1.5 of yellow] (pink) {};
        \node[node, color=c, below=1.5 of yellow] (red) {};
        \node[node, color=i, below=1.5 of pink] (blue) {};

        \draw[edg] (yellow) -- (pink);
        \draw[edg] (yellow) -- (blue);
        \draw[edg] (red) -- (pink);
        \draw[edg] (red) -- (blue);
        \draw[edg] (pink) -- (blue);
        
        \node[label,above=0.05 of yellow] {$\mathbf a$};
        \node[label,below=0.05 of red]    {$\mathbf c$};
        \node[label,above=0.05 of pink]   {$\mathbf b$};
        \node[label,below=0.05 of blue]   {$\mathbf d$};

        \coordinate[right=0.75 of red] (m0);
        \node[label,below=0.75 of m0]   {(a)};

    \end{tikzpicture}
    \hspace{1em}
    \begin{tikzpicture}[scale=0.9, every node/.append style={transform
        shape},node/.style={fill,circle,inner sep=1,outer sep=1,color=black},
        nsmall/.style={fill,circle,inner sep=1,outer sep=1,color=black},
        arrow/.style={>=stealth,line width=1.5,color=gray!50},
        label/.style={font=\scriptsize\sffamily\bfseries,text
        depth=0pt,color=black}, node distance=1,auto,every loop/.style={}]
        \node[node] (g1) {};
        \node[node, right=0.5 of g1] (g2) {};
        \node[node, right=0.5 of g2] (g3) {};
        \node[node, right=0.5 of g3] (g4) {};
        \node[node, right=0.5 of g4] (g5) {};
        \node[node, right=0.5 of g5] (g6) {};
        \node[node, right=0.5 of g6] (g7) {};
        \node[node, right=0.5 of g7] (g8) {};
        \node[node, right=0.5 of g8] (g9) {};
        \node[node, right=0.5 of g9] (g10) {};
        \path[arrow,->,out=90,in=90,looseness=1.7,color=a] (g1) edge
        node[label, midway]{$\mathbf a$} (g2);
        \path[arrow,->,out=-90,in=-90,looseness=1.7,color=a] (g2) edge
        node[label, midway]{$\mathbf{\bar a}$} (g1);
        \path[arrow,->,out=90,in=90,looseness=1.7,color=l] (g3) edge
        node[label, midway]{$\mathbf b$} (g4);
        \path[arrow,->,out=-90,in=-90,looseness=1.7,color=l] (g4) edge
        node[label, midway]{$\mathbf{\bar b}$} (g3);
        \path[arrow,->,out=90,in=90,looseness=1.7,color=c] (g5) edge
        node[label, midway]{$\mathbf c$} (g6);
        \path[arrow,->,out=-90,in=-90,looseness=1.7,color=c] (g6) edge
        node[label, midway]{$\mathbf{\bar c}$} (g5);
        \path[arrow,->,out=90,in=90,looseness=1.7,color=i] (g7) edge
        node[label, midway]{$\mathbf d$} (g8);
        \path[arrow,->,out=-90,in=-90,looseness=1.7,color=i] (g8) edge
        node[label, midway]{$\mathbf{\bar d}$} (g7);
        \path[arrow,->,out=90,in=90,looseness=1.7] (g9) edge
        node[label, midway]{ {\boldmath $\ast$}} (g10);
        \path[arrow,->,out=-90,in=-90,looseness=1.7] (g10) edge
        node[label, midway]{{\boldmath $\ast$}} (g9);

        \node[node, below=1.6 of g1] (h1) {};
        \node[node, right=0.5 of h1] (h2) {};
        \node[node, right=0.5 of h2] (h3) {};
        \node[node, right=0.5 of h3] (h4) {};
        \node[node, right=0.5 of h4] (h5) {};
        \node[node, right=0.5 of h5] (h6) {};
        \node[node, right=0.5 of h6] (h7) {};
        \node[node, right=0.5 of h7] (h8) {};
        \node[node, right=0.5 of h8] (h9) {};
        \node[node, right=0.5 of h9] (h10) {};
        \node[node, right=0.5 of h10] (h11) {};
        \node[node, right=0.5 of h11] (h12) {};
        \path[arrow,->,out=90,in=90,looseness=1.7,color=a] (h1) edge
        node[label, midway]{$\mathbf{ab}$}(h2);
        \draw[arrow,->,out=90,in=90,looseness=1.7,dash pattern= on 5pt off
        5pt,dash phase=4pt,color=l] (h1) edge (h2);
        \path[arrow,->,out=-90,in=-90,looseness=1.7] (h2) edge node[label,
        midway]{{\boldmath $\emptyset$}} (h1);
        \path[arrow,->,out=90,in=90,looseness=1.7,color=a] (h3) edge
        node[label, midway]{$\mathbf{\bar ad}$} (h4);
        \draw[arrow,->,out=90,in=90,looseness=1.7,dash pattern= on 5pt off
        5pt,dash phase=4pt,color=i] (h3) edge (h4);
        \path[arrow,->,out=-90,in=-90,looseness=1.7] (h4) edge node[label,
        midway]{{\boldmath $\emptyset$}} (h3);
        \path[arrow,->,out=90,in=90,looseness=1.7,color=c] (h5) edge
        node[label, midway]{$\mathbf{c \bar d}$} (h6);
        \draw[arrow,->,out=90,in=90,looseness=1.7,dash pattern= on 5pt off
        5pt,dash phase=4pt,color=i] (h5) edge (h6);
        \path[arrow,->,out=-90,in=-90,looseness=1.7] (h6) edge node[label,
        midway]{{\boldmath $\emptyset$}} (h5);
        \path[arrow,->,out=90,in=90,looseness=1.7,color=l] (h7) edge
        node[label, midway]{$\mathbf{\bar b}$} (h8);
        \path[arrow,->,out=-90,in=-90,looseness=1.7] (h8) edge node[label,
        midway]{{\boldmath $\emptyset$}} (h7);
        \path[arrow,->,out=90,in=90,looseness=1.7,color=c] (h9) edge
        node[label, midway]{$\mathbf{\bar c}$} (h10);
        \path[arrow,->,out=-90,in=-90,looseness=1.7] (h10) edge node[label,
        midway]{{\boldmath $\emptyset$}} (h9);
        \path[arrow,->,out=90,in=90,looseness=1.7] (h11) edge
        node[label, midway]{ {\boldmath $\ast$}} (h12);
        \path[arrow,->,out=-90,in=-90,looseness=1.7] (h12) edge
        node[label, midway]{{\boldmath $\ast$}} (h11);

        \node[node, below=1.6 of h1] (i1) {};
        \node[node, right=0.5 of i1] (i2) {};
        \node[node, right=0.5 of i2] (i3) {};
        \node[node, right=0.5 of i3] (i4) {};
        \node[node, right=0.5 of i4] (i5) {};
        \node[node, right=0.5 of i5] (i6) {};
        \node[node, right=0.5 of i6] (i7) {};
        \node[node, right=0.5 of i7] (i8) {};
        \node[node, right=0.5 of i8] (i9) {};
        \node[node, right=0.5 of i9] (i10) {};
        \node[node, right=0.5 of i10] (i11) {};
        \node[node, right=0.5 of i11] (i12) {};
        \node[node, right=0.5 of i12] (i13) {};
        \node[node, right=0.5 of i13] (i14) {};
        \path[arrow,->,out=90,in=90,looseness=1.7,color=l] (i1) edge
        node[label, midway]{$\mathbf{bd}$}(i2);
        \draw[arrow,->,out=90,in=90,looseness=1.7,dash pattern= on 5pt off
        5pt,dash phase=4pt,color=i] (i1) edge (i2);
        \path[arrow,->,out=-90,in=-90,looseness=1.7] (i2) edge node[label,
        midway]{{\boldmath $\emptyset$}} (i1);
        \path[arrow,->,out=90,in=90,looseness=1.7,color=l] (i3) edge
        node[label, midway]{$\mathbf{\bar bc}$} (i4);
        \draw[arrow,->,out=90,in=90,looseness=1.7,dash pattern= on 5pt off
        5pt,dash phase=4pt,color=c] (i3) edge (i4);
        \path[arrow,->,out=-90,in=-90,looseness=1.7] (i4) edge node[label,
        midway]{{\boldmath $\emptyset$}} (i3);
        \path[arrow,->,out=90,in=90,looseness=1.7,color=a] (i5) edge
        node[label, midway]{$\mathbf{a}$} (i6);
        \path[arrow,->,out=-90,in=-90,looseness=1.7] (i6) edge node[label,
        midway]{{\boldmath $\emptyset$}} (i5);
        \path[arrow,->,out=90,in=90,looseness=1.7,color=a] (i7) edge
        node[label, midway]{$\mathbf{\bar a}$} (i8);
        \path[arrow,->,out=-90,in=-90,looseness=1.7] (i8) edge node[label,
        midway]{{\boldmath $\emptyset$}} (i7);
        \path[arrow,->,out=90,in=90,looseness=1.7,color=c] (i9) edge
        node[label, midway]{$\mathbf{\bar c}$} (i10);
        \path[arrow,->,out=-90,in=-90,looseness=1.7] (i10) edge node[label,
        midway]{{\boldmath $\emptyset$}} (i9);
        \path[arrow,->,out=90,in=90,looseness=1.7,color=i] (i11) edge
        node[label, midway]{$\mathbf{\bar d}$} (i12);
        \path[arrow,->,out=-90,in=-90,looseness=1.7] (i12) edge node[label,
        midway]{{\boldmath $\emptyset$}} (i11);
        \path[arrow,->,out=90,in=90,looseness=1.7] (i13) edge
        node[label, midway]{ {\boldmath $\ast$}} (i14);
        \path[arrow,->,out=-90,in=-90,looseness=1.7] (i14) edge
        node[label, midway]{{\boldmath $\ast$}} (i13);
        
        \coordinate[left=0.39 of g1] (g0);
        \coordinate[left=0.39 of h1] (h0);
        \coordinate[left=0.39 of i1] (i0);

        \node[anchor=east] at (g0) {$G$};
        \node[anchor=east] at (h0) {$H$};
        \node[anchor=east] at (i0) {$I$};

        \coordinate[right=0.25 of i7] (m0);
        \node[label,below=1 of m0]   {(b)};

    \end{tikzpicture}
    \end{center}
    \caption{(a) A simple graph bounded by degree three and (b) a corresponding
    FF-Median instance constructed with transformation scheme
    \textbf{R}.}\label{fig:reduction}
\end{figure}
\begin{example}
    Figure~\ref{fig:reduction}~(b) shows a FF-Median instance constructed with
    transformation scheme \textbf{R} from the simple graph depicted in
    Figure~\ref{fig:reduction}~(a). Gene similarities between genes are not
    shown, but can be derived from the genes' labeling. 
\end{example}

We structure our proof that the presented transformation is in fact a valid
mapping of an MAX~IS-3 instance to an instance of FF-Median into three
different lemmas: 

\begin{lemma}\label{lem:med_genes} Given a median $M$ of FF-Median instance
    $\mathbf R(\Lambda) = (G, H, I, \sigma)$, (1) for each median gene $(g, h,
    i) \in \mathcal C(M)$ where $g$, $h$, or $i$ are associated with vertices in
    $V(\Lambda)$ holds $\xi(g) = \xi(h) \cap \xi(i) = v$, $v \in V(\Lambda)$;
    (2) there exist at most two median genes whose corresponding extant genes
    are not associated to any vertex in $V(\Lambda)$.
\end{lemma}

\proof 
Assume for contradiction that claim (1) does not hold. Then either $\xi(g) \neq
\xi(h) \cap \xi(i)$, or $\xi(h) \cap \xi(i) = \emptyset$, both of which violate
the constraint of establishing gene similarities between associated genes that
is given in step 5. For claim (2), observe that the only unassociated genes in
genome $G$ are gene $g_\ast$ and $\bar g_\ast$ introduced in step 6, limiting
the overall number of unassociated genes in any median $M$. \qed

\begin{lemma}\label{lem:med_adj} The conserved adjacency set of any median $M$
    of FF-Median instance $\mathbf R(\Lambda) = (G, H, I, \sigma)$ is of the
    form $\mathcal A(M) \cap \mathcal A_\Yup^C = \mathcal A_\Yup^G(M) \cup \{
    \{m_\ast^\textnormal{h}, \overline m_\ast^\textnormal{t}\}, \{\overline
    m_\ast^\textnormal{h}, m_\ast^\textnormal{t}\} \}$, where the extant genes
    corresponding to $m_\ast$ and $\overline m_\ast$ are all unassociated genes
    of type ``$\ast$'' and $\mathcal A(M)_\Yup^G \subseteq \left\{
    \{m_1^\textnormal{h}, m_2^\textnormal{t}\} \in A_\Yup^C~|~\xi(\pi_G(m_1)) =
    \xi(\pi_G(m_2))\right\}$. 
\end{lemma}

\proof Observe that both candidate median adjacencies $a_\ast =
\{m_\ast^\text{h}, \overline m_\ast^\text{t}\}$ and $\bar a_\ast = \{\overline
m_\ast^\text{h}, m_\ast^\text{t}\}$ are conserved in all three genomes, whereas
all other conserved candidate adjacencies between associated and unassociated
genes can be at most conserved in $H$ and $I$. 
Establishing adjacencies $a_\ast,\bar a_\ast$ gives rise to a cumulative
adjacency score of $6$.  Conversely, up to $4$ non-conflicting adjacencies
between associated and unassociated genes can be established that are conserved
in both genomes $H$ and $I$. However, since such adjacencies are only conserved
between unassociated genes of type ``$\emptyset$'' whose gene similarities are
set to $\frac{1}{4}$, the best cumulative adjacency score can not exceed $4$.
Thus, adjacencies $a_\ast, \bar a_\ast$ must be contained in any median.
Further, because of this and the fact that in both genomes $H$ and $I$, each
gene associated with vertices of $V(\Lambda)$ is only adjacent to an
unassociated gene, $M$ cannot contain adjacencies that are conserved in extant
genomes other than genome $G$, which are the adjacencies of each gene pair
$(g_v, \bar g_v)$ associated with the same vertex $v \in V(\Lambda)$.  \qed

\begin{lemma}\label{lem:conflict} Given FF-median instance $\mathbf R(\Lambda) =
    (G, H, I, \sigma)$, let $m_u, m_v$ be any pair of candidate median
    adjacencies of $\mathcal A_\Yup$ whose corresponding extant genes are
    associated to vertices $u, v \in V(\Lambda)$, then $m_u, m_v$ are
    conflicting if and only if $(u, v) \in E$. 
\end{lemma}

\proof By construction in step 5 of transformation scheme \textbf{R}, each
vertex $v \in V$ is associated with exactly two candidate median genes $m_v =
(g, h, i), \overline m_v = (\bar g, \bar h, \bar i)$, $m_v, \overline m_v \in
\Sigma_\Yup$, such that $\xi(g) = \xi(h) \cap \xi(i) = v $ and $\xi(\bar{g}) =
\xi(\bar h) \cap \xi(\bar i) = v$.  Further, let $u$ be another vertex of
$V(\Lambda)$, such that $(u, v) \in E(\Lambda)$, and $m_u, \overline m_u$ are
its two corresponding candidate median genes. Then, by construction in step 2,
there exists exactly one extant gene $x$ with $\xi(x) = uv$ (which, by
assignment in step 3, is either contained in genome $H$ or $I$).  Consequently,
either $m_u$ is in conflict with $m_v$, or $\overline m_u$ with $\overline m_v$,
or $\overline m_u$ with $m_v$, or $m_u$ with $\overline m_v$.
Recall that by construction in step 2 in \textbf{R} and by
Lemma~\ref{lem:med_adj}, $m_u, \overline m_u$ and $m_v, \overline m_v$ form
conserved candidate adjacencies $\{m_u^\text{h}, \overline m_u^\text{t}\}$,
$\{\overline m_u^\text{h}, m_u^\text{t}\}$ and $\{m_v^\text{h}, \overline
m_v^\text{t}\}$, $\{\overline m_v^\text{h}, m_v^\text{t}\}$, respectively.
Clearly, independent of which of the candidate median gene pairs of $u$ and $v$
are in conflict, both pairs of candidate median adjacencies are in conflict
with each other.

Now, let $u, v$ be two vertices of $V(\Lambda)$ such that edge $(u, v) \not \in
E(\Lambda)$, then there exists no gene $x$ in extant genomes $H$ and $I$ with
$\xi(x) = uv$. Even more, due to Lemma~\ref{lem:med_genes}, there cannot exist
a candidate median gene $(g, h, i)$ with $\{u, v\} \subseteq \xi(g) \cup \xi(h)
\cup \xi(i)$. Thus, the candidate median genes of $u$ and $v$ are not
conflicting and neither are their corresponding candidate median adjacencies.
\qed

We proceed to show that the given transformation scheme gives rise to an
approximation preserving reduction known as \emph{L-reduction}. An L-reduction
reduces a problem $P$ to a problem $Q$ by means of two polynomial-time
computable transformation functions: A function $f: P \to Q' \subseteq Q$ that
maps each instance of $P$ onto an instance of $Q$, herein represented by
transformation scheme \textbf{R}, and a function $g: Q' \to P$ to transform any
feasible solution of an instance in $Q'$ to a feasible solution of an instance
of $P$. Here, a \emph{feasible} solution means any -- not necessarily
\emph{optimal} -- solution that obeys the problem's constraints. A feasible
solution of FF-Median instance $(G, H, I, \sigma)$ is an \emph{ancestral genome}
$X$ where $\mathcal C(X) \subseteq \Sigma_\Yup$ and $\mathcal A(X) \subseteq
\mathcal A_\Yup$ such that $\mathcal A(X)$ is conflict-free. We give the
following transformation scheme to map ancestral genomes of an FF-Median
instance to solutions of an MAX~IS-3 instance: 

\bigskip 

\noindent Transformation \textbf{S}: Given any ancestral genome $X$ of $\mathbf
R(\Lambda)$, return $\{\xi(\pi_G(m_1))~|~\allowbreak\{m_1^a, m_2^b\} \in \mathcal A(X):
\mathbb I_G(\pi_G(m_1)^a, \pi_G(m_2)^b) = 1 \text{ and } \xi(\pi_G(m_1))
\neq \emptyset \}$. 

\bigskip

\noindent We define score function $s_\Yup(X) \equiv \frac{1}{2}\mathcal
F_\Yup(X) - 3$ of an ancestral genome $X$. For $(\mathbf R, \mathbf S)$ to be
an L-reduction the following two properties must hold for any given MAX~IS-3
instance $(\Lambda, l)$: (1) There is some constant $\alpha$ such that for any
median $M$ of the transformed FF-Median instance $\mathbf R(\Lambda)$ holds
$s_\Yup(M) \leq \alpha \cdot l$; (2) There is some constant $\beta$ such that
for any ancestral genome $X$ of $\mathbf R(\Lambda)$ holds $l - |\mathbf
S(X)| \leq \beta \cdot |s_\Yup(M) - s_\Yup(X)|$. We proceed to proof the
following lemma:

\begin{lemma} $(\mathbf R, \mathbf S)$ is an L-reduction of problem MAX~IS-3 to
    problem FF-Median with $\alpha = \beta = 1$.
\end{lemma}

\proof 
For any median $M$ of FF-Median instance $\mathbf R(\Lambda)$, the
number of conserved median adjacencies with correspondence to the same vertex
of $\Lambda$ is two, giving rise a cumulative adjacency score of two.  From
Lemmata~\ref{lem:med_adj} and \ref{lem:conflict} immediately follows that any
ancestral genome of $\mathbf R(\Lambda)$ that maximizes the number of conserved
adjacencies also maximizes the number of independent vertices in $\Lambda$.
Recall that the two conserved adjacencies between unassociated genes of type
``$\ast$'' (which are part of all medians) give rise to a cumulative adjacency
score of $6$, we conclude that $|\mathcal A(M) \cap \mathcal A_\Yup^C| - 2 =
\frac{1}{2} \mathcal F_\Yup(M) - 3 = s_\Yup(M) = l$, thus $\alpha = 1$.

Because $l = s_\lambda(M)$, it remains to show that $l-|S(X)| \leq \beta
|l-s_\Yup(X)|$.  In a \emph{sub-optimal} ancestral genome of $\mathbf
R(\Lambda)$, median genes with no association to vertices of $\Lambda$ can also
contain extant genes of type ``$\emptyset$''. These unassociated median genes
can form ``mixed'' conserved adjacencies with genes that are associated with
vertices of $\Lambda$.  Such mixed conserved adjacencies have no correspondence
to vertices in $\Lambda$ and do not contribute to the transformed solution
$\mathbf S(X)$ of an ancestral genome $X$. Yet, as mentioned earlier, the
cumulative adjacency score of all mixed conserved adjacencies can not not
exceed $4$. Therefore it holds that $|S(X)| \geq s_\Yup(X)$ and we conclude
$\beta = 1$. \qed


\section{Speeding up the search for a median}\label{app:eg}

Proof of Lemma~\ref{lem:ICF-SEG}: 

\proof 
Given an IC-free segment $\mathcal S= \{ m_1,\ldots,m_k\}$ of an FF-Median
instance $(G, H, I, \sigma)$. Let $\mathcal M$ be a conflict-free matching in
graph $\Gamma'(\mathcal S)$. Because $\mathcal M$ is conflict-free and $\mathcal
S$ contiguous in all three extant genomes, $M$ must contain all candidate median
genes of $S$. Now, let $M$ be a median such that $\mathcal S \not\subseteq
\mathcal C(M')$. Further, let $C(m)$ be the set of candidate median genes that
are involved in a conflict with with a given median gene $m$ of $\mathcal S$ and
$X = \mathcal C(M') \cap (\bigcup_{m \in \mathcal S} C(m) \cup \mathcal S)$.
Clearly, $X \neq \emptyset$ and for the contribution $\mathcal F_\Yup(X)$ must
hold $\mathcal F_\Yup(X) \geq \mathcal F_\Yup(\mathcal S)$, otherwise $M'$ is
not optimal since it is straightforward to construct a median higher score which
includes $\mathcal S$. Clearly, the contribution $\mathcal F(X)$ to the median
is bounded by $\max(\{\sum_{m'\in C'}\Delta(m')~|~C' \subseteq C(m) :~ C' \text{
is conflict-free}\}) + \mathcal F_\Yup(\mathcal S)$. But since $\mathcal S$
gives rise to a conflict-free matching with maximum score, also median $M''$
with $\mathcal C(M'') = (\mathcal C(M') \setminus X) \cup \mathcal C(\mathcal
S)$ and $\mathcal A(M'') = (\mathcal A(M') \setminus \mathcal A(X)) \cup
\mathcal A(S))$ must be an (optimal) median. \qed

\section{Simulated sequence evolution with ALF}\label{app:simulations}
\begin{table}[h!]
     \centering
     {\footnotesize
     \begin{tabular}{rcrrrr}
         \toprule
         \textbf{PAM} & \textbf{Genome}&\textbf{Inversions}&\textbf{Transpositions}&\textbf{Duplications}&\textbf{Losses}\\
         \midrule
         \multirow{3}{*}{10} &$G$      &8.7       &6.1      &7.3      &6.9   \\ 
                             &$H$      &7.3       &4.5      &6.3      &5.4   \\ 
                             &$I$      &8.5       &6.6      &10.4     &5.6 \medskip\\ 
         \multirow{3}{*}{30} &$G$      &24.5      &16.9     &21.0     &22.7 \\
                             &$H$      &23.4      &19.8     &20.6     &18.4 \\
                             &$I$      &25.5      &17.2     &17.5     &20.9\medskip\\
         \multirow{3}{*}{50} &$G$      &39.9      &27.8     &32.4     &36.7 \\
                             &$H$      &41.8      &31.8     &31.0     &31.7 \\
                             &$I$      &43.2      &30.0     &28.7     &39.7\medskip\\
         \multirow{3}{*}{70} &$G$      &58.6      &42.3     &41.1     &39.2   \\
                             &$H$      &57.0      &43.6     &46.3     &45.1   \\
                             &$I$      &60.4      &41.4     &40.7     &39.1\medskip\\
         \multirow{3}{*}{90} &$G$      &75.0      &54.5     &53.1     &64.2 \\
                             &$H$      &69.9      &50.5     &54.1     &65.0  \\
                             &$I$      &75.2      &55.5     &60.3     &58.5\medskip\\
         \multirow{3}{*}{110}&$G$      &96.3      &69.4     &67.0     &74.6 \\
                             &$H$      &90.6      &64.2     &62.5     &70.9 \\
                             &$I$      &90.2      &68.5     &62.6     &61.2\medskip\\
         \multirow{3}{*}{130}&$G$      &105.7     &76.3     &74.4     &81.0   \\
                             &$H$      &108.7     &78.2     &79.6     &82.8 \\
                             &$I$      &110.8     &73.6     &73.9     &77.3\medskip\\
         \bottomrule
     \end{tabular}
     }
     \caption{Average benchmark data of seven evolutionary distances, each
     comprising ten genomic datasets generated by
     ALF~\cite{Dalquen:2012dx}.}\label{tab:median_data_info}
\end{table}

\begin{table}[tb]
    \centering
    {\scriptsize
    \begin{tabular}{rll}
        \toprule
        \textbf{Parameter name} & \textbf{Value}\\
        \midrule
        \multicolumn{3}{l}{\textit{sequence evolution}}\\
        \midrule
        substitution model&\multicolumn{2}{l}{WAG (amino acid substitution
        model)}\\
        insertion and deletion&Zipfian distribution&exponent $c=1.8214$\\
        & insertion rate & $0.0003$\\
        & maximum insertion length & $50$\\
        rate variation among sites & $\Gamma$-distribution & shape parameter $a=1$ \\
        &number of classes&$5$\\
        &rate of invariable sites & $0.01$\\
        \midrule
        \multicolumn{3}{l}{\textit{genome rearrangement}}\\
        \midrule
        inversion&rate&$0.0004$\\
        &maximum inversion length&$100$\\
        transposition&rate&$0.0002$\\
        &maximum transposition length&$100$\\
        &rate of inverted transposition&$0.1$\\
        \midrule
        \multicolumn{3}{l}{\textit{gene family evolution}}\\
        \midrule
        gene duplication&rate&$0.0001$\\
        &max. no. of genes involved in one dupl.&$5$\\
        &probability of transposition after dupl.&$0.5$\\
        &fission/fusion after duplication&$0.1$\\
        &probability of rate change&$0.2$\\
        &rate change factor&$0.9$\\
        &probability of temporary rate change (duplicate)&$0.5$\\
        &temporary rate change factor (duplicate)&$1.5$\\
        &life of rate change (duplicate)&$10$ PAM\\
        &probability of temporary rate change (orig+duplicate)&$0.3$\\
        &temporary rate change factor (orig+duplicate)&$1.2$\\
        &life of rate change (orig+duplicate)&$10$ PAM\\
        gene loss&rate&$0.0001$\\
        &maximum length of gene loss&$5$\\
        gene fission/fusion&rate&$0.0$\\
        &maximum number of fused genes&$-$\\

        \bottomrule

    \end{tabular}
    }
    \caption{Parameter settings for simulations generated by
    ALF~\cite{Dalquen:2012dx}.}\label{tab:alf_params}
\end{table}

\FloatBarrier
\newpage
\section{Real genomes dataset}\label{app:real_genomes}

\begin{table}[h!]
    \centering
    {\scriptsize
    \begin{tabular}{ll}
        \toprule
        \textbf{Genbank ID} & \textbf{Name}\\
        \midrule
        U00096.3 & Escherichia coli str. K-12 substr. MG1655, complete genome\\
        AE004439.1 & Pasteurella multocida subsp. multocida str. Pm70, complete genome\\
        AE016853.1 & Pseudomonas syringae pv. tomato str. DC3000, complete genome\\
        AM039952.1 & Xanthomonas campestris pv. vesicatoria complete genome\\
        CP000266.1 & Shigella flexneri 5 str. 8401, complete genome\\
        CP000305.1 & Yersinia pestis Nepal516, complete genome\\
        CP000569.1 & Actinobacillus pleuropneumoniae L20 serotype 5b complete genome\\
        CP000744.1 & Pseudomonas aeruginosa PA7, complete genome\\
        CP000766.3 & Rickettsia rickettsii str. Iowa, complete genome\\
        CP000950.1 & Yersinia pseudotuberculosis YPIII, complete genome\\
        CP001120.1 & Salmonella enterica subsp. enterica serovar Heidelberg str. SL476, complete genome\\
        CP001172.1 & Acinetobacter baumannii AB307-0294, complete genome\\
        CP001363.1 & Salmonella enterica subsp. enterica serovar Typhimurium str. 14028S, complete genome\\
        FM180568.1 & Escherichia coli 0127:H6 E2348/69 complete genome, strain E2348/69\\
        CP002086.1 & Nitrosococcus watsoni C-113, complete genome\\
        \bottomrule

        \end{tabular}
        }
        \caption{Dataset of genomes used in comparison with the OMA
        database.}\label{tab:oma_genomes}
    \end{table}

\end{document}